\documentclass[traditabstract]{aa} 
                                   
\usepackage{graphicx}

\usepackage{txfonts}
\usepackage{natbib}
\bibpunct{(}{)}{;}{a}{}{,} 

\begin{document}
   \title{Photophoretic separation of metals and silicates: the formation of Mercury like planets and metal depletion in chondrites}

   \author{Gerhard Wurm,
          \inst{1} \and
          Mario Trieloff
          \inst{2}
          \and
          Heike Rauer\inst{3,4}
          }

   \institute{Fakult\"at f\"ur Physik, Universit\"at Duisburg-Essen,
              Lotharstr. 1, D-47057 Duisburg \and Institut f\"ur Geowissenschaften, Universit\"at Heidelberg, Im Neuenheimer Feld 234-236, D-69120 Heidelberg \and 
              Institut f\"ur Planetenforschung, Extrasolare Planeten und Atmosph\"aren, Deutsches Zentrum f\"ur Luft- und Raumfahrt (DLR), Rutherfordstrasse 2, D-12489 Berlin \and
              Zentrum f\"ur Astronomie und Astrophysik, Technische Universit\"at Berlin, Marchstrasse 6, D-10587 Berlin\\
              \email{gerhard.wurm@uni-due.de}
             }

   \date{Received ; accepted }
 
  \abstract
   {Mercury's high uncompressed mass density suggests that the planet is largely composed of iron, either bound within metal (mainly Fe-Ni), or iron sulfide. Recent results from the MESSENGER mission to Mercury imply a 
low temperature history of the planet which questions the standard formation models of impact mantle stripping or evaporation to explain the high metal content. 
Like Mercury, the two smallest extrasolar rocky planets with mass and size determination, CoRoT-7b and Kepler-10b,  were  found to be of high density. As they orbit close to their host stars this indicates that iron rich {\it inner} planets might not be a nuisance of the solar system but be part of a general scheme of planet formation. From undifferentiated chondrites it is 
also known that the metal to silicate ratio is highly variable which must be ascribed to pre-planetary fractionation processes. Due to this fractionation most chondritic parent bodies -- most of them originated in the asteroid belt -- are depleted in iron relative to average solar system abundances.
The astrophysical processes leading to metal silicate fractionation in the solar nebula are essentially unknown. 
Here, we consider photophoretic forces. 
As these forces particularly act on irradiated solids, they might play a significant role for the composition of planetesimals forming at the inner edge of protoplanetary discs. 
Photophoresis can separate high thermal conductivity materials (iron) from lower thermal conductivity solids (silicate). We suggest that the silicates are preferentially pushed into the optical thick disk. Subsequent planetesimal formation at the edge moving outwards leads to metal rich planetesimals close to the star and metal depleted planetesimals further out in the nebula.}

\titlerunning{Photophoretic separation of metals and silicates}
   \maketitle

\section{Introduction}

It is long known that Mercury is a rather dense planet. With an average uncompressed density of  $5.3 \rm g/cm^3$ its average density is much higher than Earth's uncompressed density of $4.4 \rm g/cm^3$ or the uncompressed density of Mars of $3.8 \rm g/cm^3$ \citep{spohn2001}.  This suggests that Mercury has an iron rich core which is much larger than the core of other terrestrial planets especially in relation to the small size of the planet. There is a general trend in the inner solar system that the uncompressed density of the planets or asteroids decreases  with distance to the sun. 

\begin{figure}[h]
\includegraphics[width=\columnwidth]{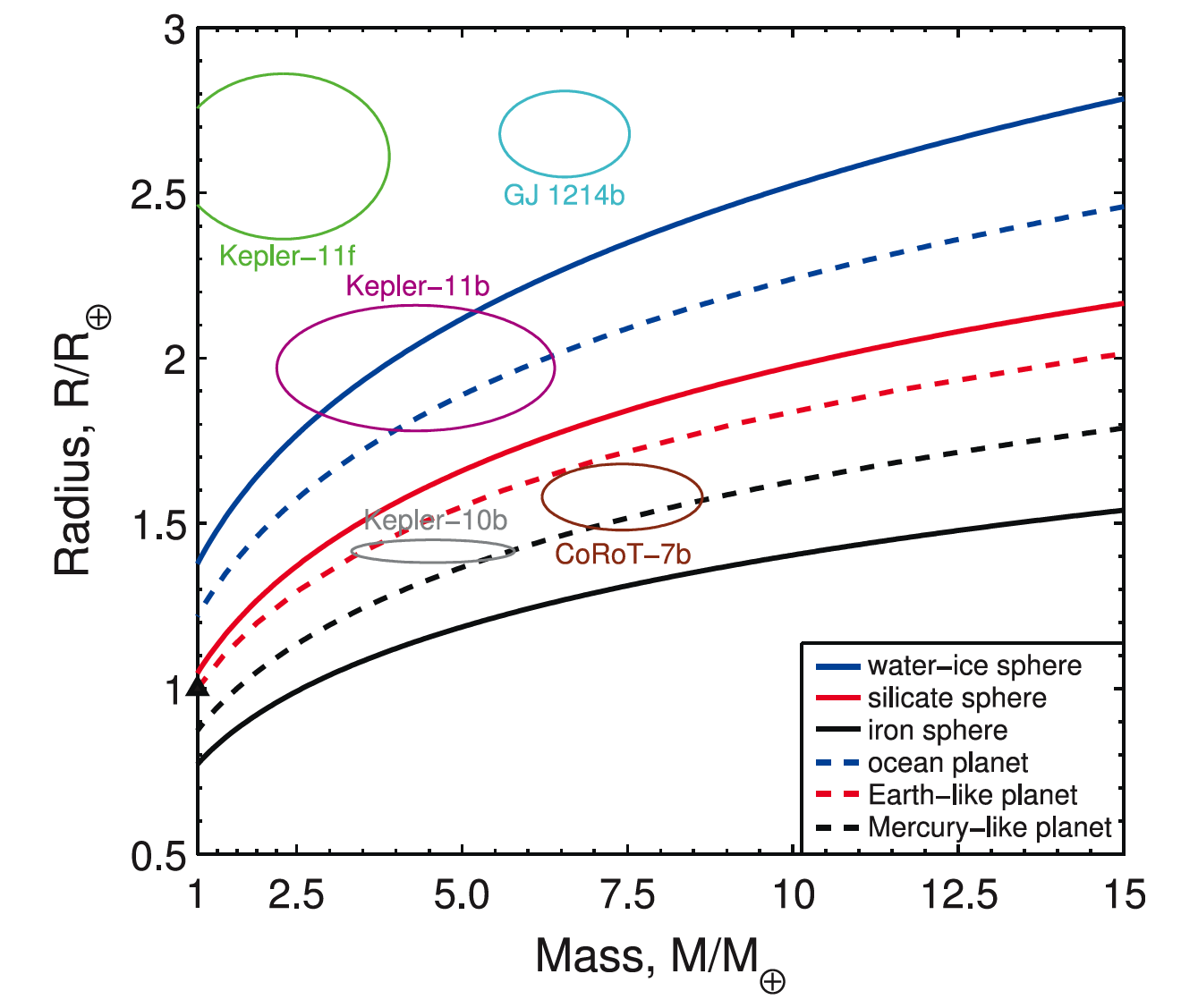}
    \caption{\label{wagner2011} Most recent models place CoRoT-7b and Kepler-10b on a mass-radius diagram close to a composition similar to Mercury (color online).  (from \citep{wagner2011})}
\end{figure}

Recently, the first extrasolar planets with masses below 10 $\rm M_{Earth}$, so-called super-Earths, have been detected (see e.g. the extrasolar planet encyclopedia for an updated list at exoplanet.eu). The smallest of these objects for which true masses and radii could be constrained are CoRoT-7b \citep{leger2009} and Kepler-10b \citep{batalha2011}. Both planets are small, low-mass objects which are most likely of rocky nature. Since both planets orbit rather faint host stars, the determination of their masses by ground-based spectroscopy has been challenging. In particular for CoRoT-7b parameters have been revised several times in the literature (e.g. see \citet{hatzes2011} for a summary). When masses and radii of rocky planets are known, these parameters can be compared to model calculations of the planet interior structure and composition, assuming hydrostatic equilibrium and equations-of-state corresponding to the modeled composition of the planets. If mass and radii constraints are sufficient, the comparison with such models allows us to constrain the interior composition of extrasolar planets. For CoRoT-7b and Kepler-10b several such  comparisons have been made. 
Models comparing with the original, meanwhile revised CoRoT-7b parameters arise with a planet lighter than Earth \citep{swift2012}. However, all authors using the most recent and commonly accepted values for CoRoT-7b (\citet{bruntt2010} for the radius and \citet{hatzes2011} for its mass) agree that this planet has a core larger than Earth and is more like a ‘super-Mercury’ planet than an Earth-twin in its interior structure, and similar results are found for Kepler-10b \citep{wagner2011, valencia2011}. This is illustrated in Fig. \ref{wagner2011}.

Both planets orbit their stars on extremely close distances, with orbital periods of less than a day. Recently, \citet{leitzinger2011} have shown that they cannot be remnants of larger gas giant or Neptune-like planets which migrated inward. It is more likely they formed as rocky planets in the inner parts of their planetary system. If this is so, then they resemble some of the structure found in the Solar System, with a large-core, terrestrial inner planet. Unfortunately, we do not have information on radii and masses for the outer planets of these systems. Therefore, whether large cores for inner terrestrial planets, in contrast to outer planets, are a common feature of planetary systems has to await future exoplanet detections, in particular in the low-mass range. Nevertheless, that the smallest rocky planets detected so far are somewhat similar to Mercury is curious. 

There have been several explanations for the high density of Mercury. A common explanation is an impact which stripped off most of its silicate mantle \citep{Benz1988}. Another idea is that the temperatures were so high in the early solar system that part of the surface evaporated \citep{cameron1985}. These models have been challenged by the MESSENGER mission which yielded surprising results that largely rule out the impact or evaporation scenario as reason for silicate depletion of Mercury. Based on the measurements of the lithophile incompatible elements K, U and Th, \citet{peplowski2011} found a striking similarity of the K/Th ratio on the surfaces of Mercury, comparable to  Mars, Venus and the Earth. Contrary to Th, K is a volatile element and should have been measurably depleted by a giant impact \citep{Benz2007}
or large-scale planetary evaporation processes, as e.g. similar to the Moon. This is, however, not the case. 

\begin{figure}[h]
\includegraphics[width=\columnwidth]{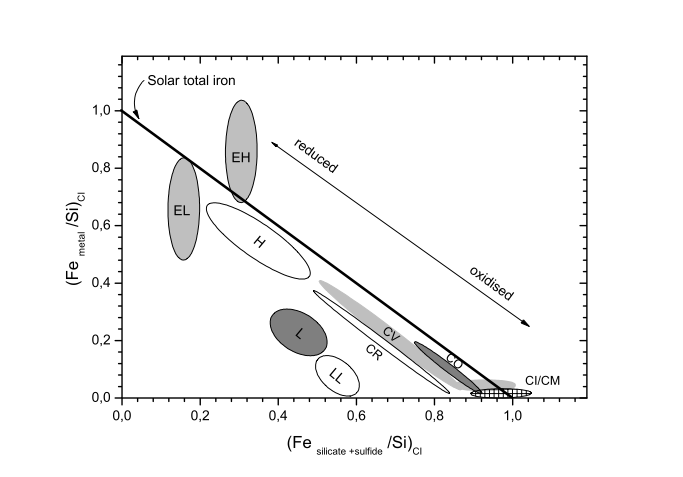}
    \caption{\label{trieloffbild} Urey-Craig diagram demonstrating
variability of total metal content as well as oxidation state among chondritic (i.e. undifferentiated) asteroids: Most are depleted in total iron, while reduced EH chondrites have an excess of Fe. (adapted from \citet{trieloff2006}) }
\end{figure}

Alternative explanations may be sought in metal-silicate fractionation processes at the preplanetary stage, and indeed primitive meteorites from undifferentiated asteroids provide ground truth that these occurred. For example, most chondritic parent bodies are depleted in total iron (either present as reduced metal or in oxidised form in silicates) relative to bulk solar system values (=CI chondrites). However, exceptions exist, e.g. EH enstatite chondrites that have an overabundance of metal as seen in Fig. \ref{trieloffbild}. 

Based on MESSENGER Gamma ray spectrometer geochemical data, enstatite chondrites were also favoured by \citet{nittler2011} as possible precursor material for Mercury, although cautioning that present day meteorite collections may not necessarily represent the full spectrum of "planetary building blocks". 
Nevertheless, these facts leave us with the lack of an explanation to fractionate metal from silicate, whether preplanetary or afterwards. The situation is even worse, as conventionally considered processes fractionating lithophile and refractory elements on the preplanetary stage \citep{lewis1973} -- if strong enough to enrich Fe in Mercury's precursor material -- would certainly decrease the K abundance stronger than indicated by MESSENGER data.

Clearly, one explanation that could reconcile all these observations -
metal rich inner planets (and only minor volatile depletion in the case of Mercury) and mostly metal poor asteroids would be desirable. 

Here we discuss photophoresis as a viable mechanism. Photophoresis essentially means that an illuminated particle embedded in a low pressure gaseous environment is subject to a force (typically) directed away from the light source. 
This mechanism does not need evaporation which 
would change the chemistry of the material. It is a pure low temperature selection effect which essentially separates
metal rich and metal poor phases by a physical process.

A prerequisite for photophoresis to work is that the part of the disk where separation occurs is optically thin and that directed (stellar) radiation illuminates the particles.
In recent years a large number of transitional disks have been observed with inner gaps which are optically thin \citep{sicilia2008, najita2007}. This
concept has been extended recently to earlier pre-transitional states with two
walls, where the space between the walls is optically thin but still filled with substantial mass \citep{Espaillat2012}. There is no doubt that photophoresis works at the inner wall
of these disks.  

As this is the first time the photophoretic separation of metals and silicates is considered we propose a rather simple model here. 
As detailed below we assume that the essential part of all planetesimals with different metal content that later form terrestrial planets and asteroids 
are generated in the thin region of an outward moving inner edge of the disk. This model  considers 3 aspects discussed in the literature with respect to planetesimal formation. It combines (1) the usual mode of initial particle growth through collision with (2) the local photophoretic transport and separation of metal and silicate grains at the edge and (3) the trigger of planetesimal formation of the most dense inner part of the edge which is metal rich. This way the first planetesimals close to the star are metal rich while the metal content decreases the further the edge is moving outwards. A high density planet would form naturally in the inner region of a protoplanetary disk, the metal poor material would be found further out up to the asteroid belt (with some metal rich asteroids being scattered from the inner system).

\section{A model for the formation of planetesimals with variable iron content}

\subsection{The inner edge}

Inner edges of protoplanetary disks show up in different settings. From the beginning of a protoplanetary disk a sublimation edge is considered close to the star, where the temperatures reach values that destroy dust particles.
Current observations and modeling shows that at later times transitional or pre-transitional disks have wall like edges \citep{Espaillat2012, Calvet2002}). 
The term {\it wall} emphasizes that the edge is rather thin, which means that it turns from optical thin to optical thick on a short length scale. 

In the later discussion we will estimate if particle transport and separation fits to the overall evolution timescales. A basic quantity for this is the particle number density. We estimate this as follows.
We take the thickness of the edge as $d=100.000 \rm km$ between optical thin gap and optical thick outer disk. We assume that each particle absorbs stellar radiation over its geometrical cross
section. In a simple estimate the disk then becomes optically thick over the given edge thickness if one particle is placed in a volume of particle cross section times edge thickness. This gives a density of 
$n=1/(\pi r^2 d)$. If we take  $r = 10 \rm \mu m$ 
as particle radius this is
$n= 32 \rm m^{-3}$.

\subsection{Evidence for irradiation}

As most part of protoplanetary discs -- where planetesimal formation usually is supposed to take place -- is optically thick, one might ask for evidence that processing in optical
thin parts took place. In fact, there is evidence that irradiated matter was incorporated into planets and their precursor material.  \citet{caffee1987} discussed the presence of individual grains in carbonaceous chondrites displaying solar flare tracks and solar wind implanted noble gases, at an abundance level that suggested irradiation contemporary with growth of the parent planetesimal. \citet{trieloff2000, trieloff2002} and \citet{ballentine2005} demonstrated the presence of solar wind implanted noble gases in  Earth's interior that must have been acquired by relatively small precursor planetesimals with sufficiently high surface to volume ratio.

This suggests that already the building stones that made up the Earth were subject to irradiation. To explain this, at least part of the precursor material that constitutes the Earth had to be exposed to solar radiation. This is certainly not possible in the dense midplane of a full protoplanetary disk at 1 AU. However, it may well be that a certain part of the accreting material was derived from locations where irradiation was feasible, e.g. from the surface of a flared disc, or -- as suggested here -- from its inner rim.

\subsection{Timescales of planet formation}

There are few constraints to pin down the formation time of planetesimals. Some aspects
of the evolution in the early solar system can be derived from meteorite studies. Chronological studies of differentiated and chondritic meteorites indicate that most meteorite parent bodies formed within 3-4 million years after the earliest cm-sized solids (Ca,Al rich inclusions), which likely formed during collapse to the first protosolar core \citep{tscharnuter2009, trieloff2006, Connelly2012}.
This indicates that large (asteroid size) bodies formed throughout the lifetime of the protoplanetary disk of a few million years.

\subsection{A model scenario}

One might combine the current ideas of timing and disk evolution in a scenario that ties the rapid formation of asteroids to the outward movement of the inner edge \citep{haack2007, moudens2011, krauss2007}. Then photophoretic sorting can act just in time on the particles before final assembly.
\citet{loesche2012} estimate that photophoretic sorting is efficient enough to separate asteroid masses on short timescales at the edge as this region
is rather dense as wall-like transition. With continuous replenishment of dust through collisions there is plenty of metal rich and metal poor dust present for a sorting process while the edge slowly moves outwards. The innermost bodies then become metal rich. The further the edge moves outward the more metal depleted is the building material for further large bodies. Eventually, the disk is dispersed and without gas no more photophoretic sorting occurs. In a final stage the large planetesimals are assembled to planets according to the usual collisional models \citep{kokubo2010}. As bodies are partially scattered throughout the solar system at these later times it can be expected that some asteroidal meteorite parent bodies have exceptional high metal content contrary to the 
general trend.

Clearly, photophoretic sorting in planet formation has some intrinsic aspects that can place it only in certain regions of protoplanetary disks. If part of the disk is optically thick, photophoresis does not work. Therefore, it might not have been important for the formation of the first larger dust assemblies in the disk.
Inside of the gas depleted gaps of transitional protoplanetary disks there might have been too little material to efficiently grow planetesimals on the spot. However, exactly in between, namely at the inner edge of protoplanetary disks the basic conditions for photophoretic sorted growth are almost perfect.  In fact, recent simulations by 
\citet{pinilla2012} show that due to 
supply by drifting particles from the outside an edge might be a preferential side to trigger planetesimal formation. 

\begin{figure}[h]
\includegraphics[width=\columnwidth]{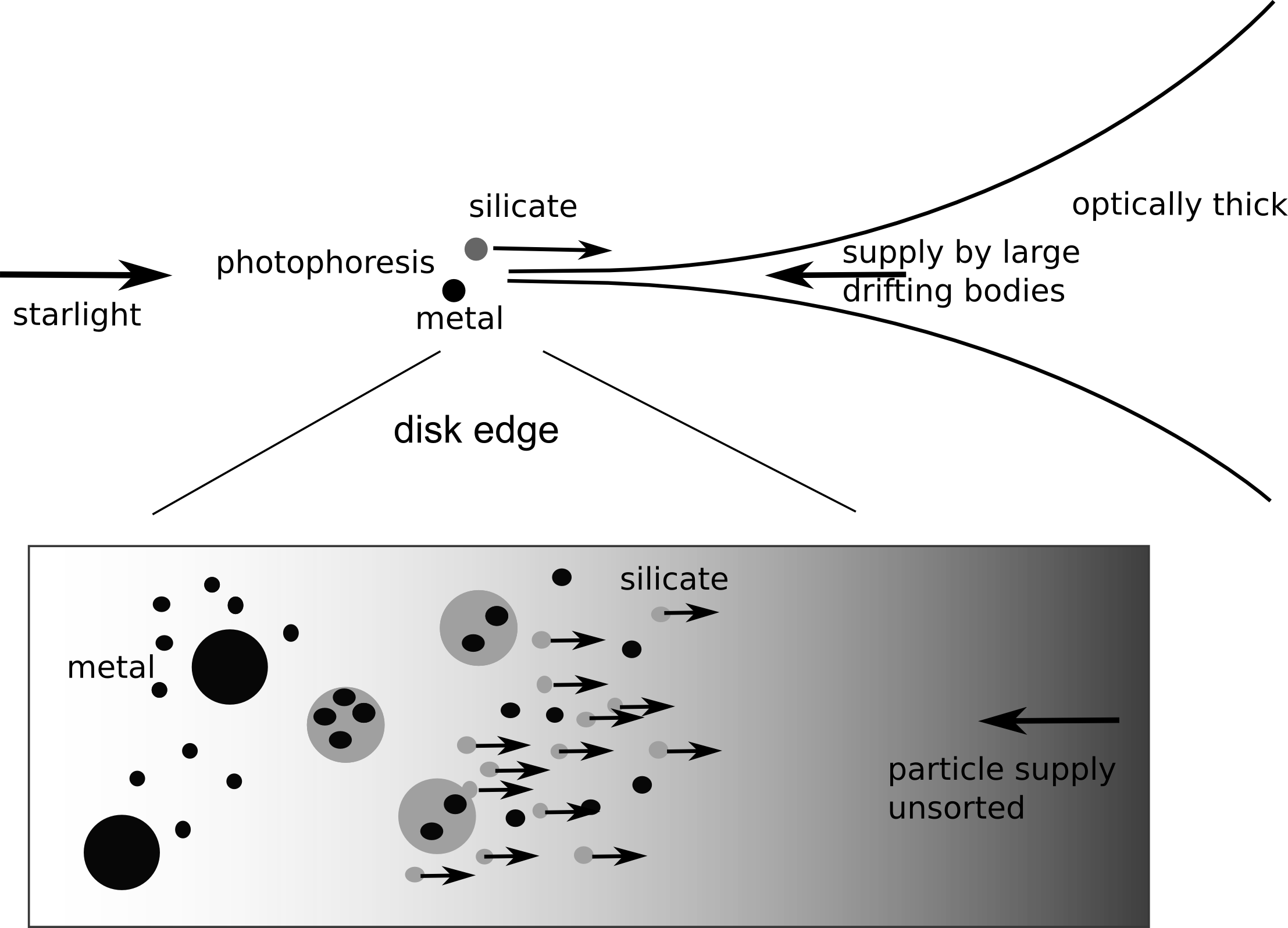}
    \caption{\label{fig:model} Basic model idea for growing metal rich inner terrestrial planets. In a 
    continuous collisional growth and fragmentation cycle silicates are transported outward while metal rich
    bodies form at the inner side of the edge.}
\end{figure}

A rough sketch of our model
can be seen in fig. \ref{fig:model}.

\subsection{Planetesimal precursors}

There has been much work on the early phases of planet formation in recent years,
namely on growth of micron-size dust particles to larger objects. From experiments
and simulations there is no doubt that at least mm-size aggregates form rapidly
\citep{blumwurm2008}. There are currently discussions about a bouncing barrier
at mm-size proposed by \citet{zsom2010}. However, collisional growth is possible
again, once cm to dm particles formed \citep{teiserwurm2009b, wurm2005, kuepper2011}. \citet{windmark2012} recently showed that this also allows the formation of planetesimals then. Another mechanism to produce planetesimals is based on 
gravitational instabilities in turbulent disks \citep{JohansenEtal2007, Chiang2010}. This depends
on a large fraction of the solid mass being present in dm-size bodies.
Therefore, there are two different populations of interest, the sub-mm particles
and the dm-bodies.

\subsection{Dust supply}

Dust particles grow rapidly to larger aggregates \citep{dullemond2005, blumwurm2008}.
Nevertheless, dust particles of micron-size are present in protoplanetary disks as long as they live, i.e. they have to be replenished for millions of years \citep{haisch2001, olofsson2010}. This has to be expected though. The collision experiments
e.g. by \citet{wurm2005} or \citet{teiserwurm2009b} show that during growth of
larger aggregates in a collision there is always a fraction of small dust particles
produced which carry the energy of the impact. There are other mechanisms
which might also produce dust particles by destroying larger objects again, i.e.
light induced erosion \citep{wurm2006b, kelling2011, wurm2010, Wurm2007, deBeule2012, Kelling2011b, Kocifaj2011} or erosion by gas drag \citep{Paraskov2006}.
Therefore, as seen in observations as well as deduced from experiments 
there is no shortcoming of individual dust grains -- metal or silicate.
The thickness of the edge, which is modeled by a wall like structure
\citep{Calvet2002} will vary according to the dust supply but as estimated
above a particle density of $10 \rm \mu m$ particles of $n = 32 \rm m^{-3}$
might be a reasonable value for further estimates here.

\subsection{Photophoretic basics}

Photophoresis is a rather old phenomenon \citep{Ehrenhaft1918}. It is known to effectively act on stratospheric and mesospheric particles in Earth's atmosphere \citep{cheremisin2011, pueschel2000}. It is technically exploited as propulsion for macroscopic objects \citep{keith2010} and it can be used for particle manipulation \citep{steinbach2004}. However, it was only introduced to astrophysical problems as physical process by \citet{krauss2005} and \citet{wurm2006}. 
Since then different applications have been exploited. Photophoresis can lead to the clearing of dust in optically thin disks \citep{krauss2007, moudens2011, herrmann2007, takeuchi2008, vonBorstel2012}. It can provide particle transport for matter to be incorporated into comets or other outer solar system bodies \citep{mousis2007, wurm2009}. It can destroy protoplanetary bodies and recycle the material \citep{Wurm2007, kelling2011, Kocifaj2011}. It can levitate small grains at the surface of protoplanetary disks \citep{wurm2009b, wurm2012} and it can provide  micron-size particles in the first place \citep{Kelling2011b}. 

The feature that is important here is the capability of photophoresis to sort material by composition. The first idea on particle sorting was originally applied to chondrules which are observed to be size sorted in meteorites (e.g. \citep{kuebler1999, scott1996}) and which can be size sorted by photophoresis \citep{wurm2006}. These studies on photophoretic separation have recently been detailed for dust mantled chondrules by \citet{loesche2012}.

\begin{figure}[h]
\includegraphics[width=\columnwidth]{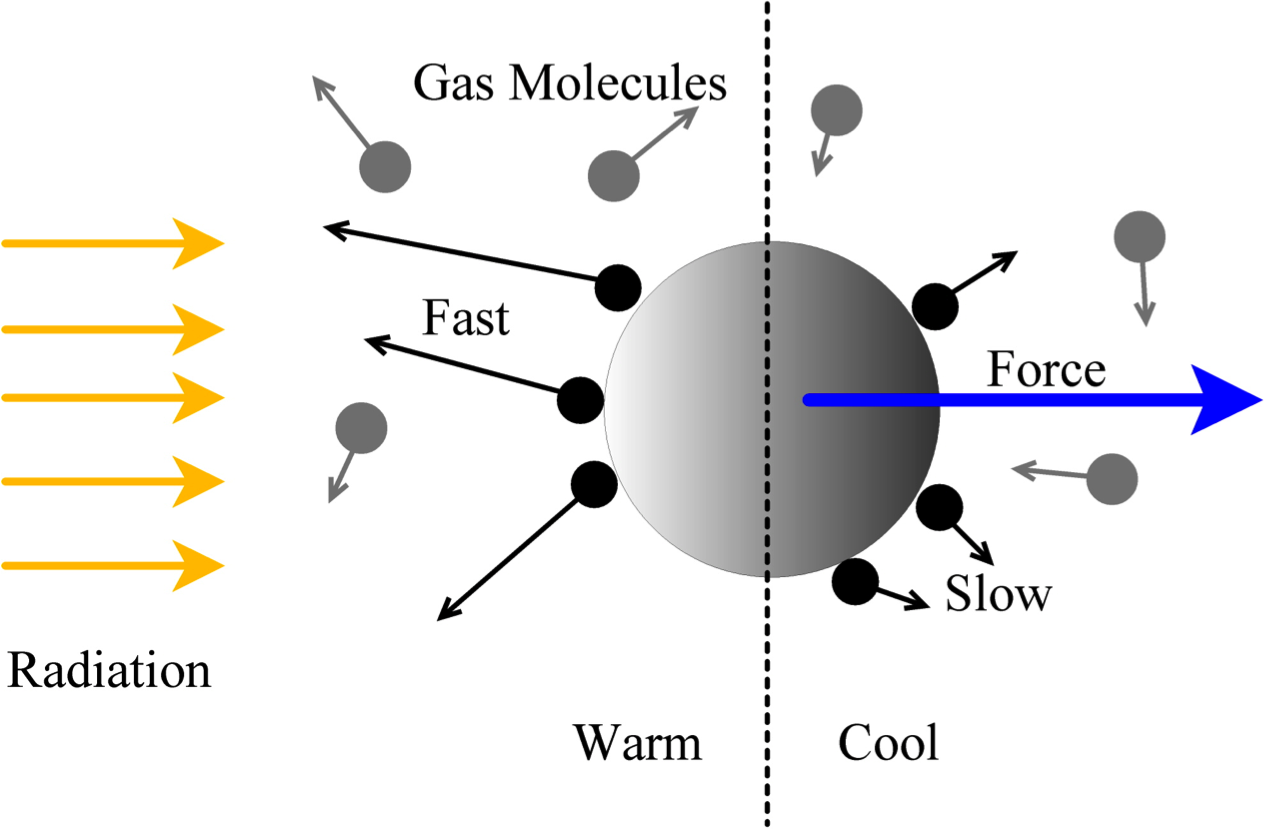}
    \caption{\label{fig:photophoresis} Photophoresis in the free molecular flow regime. Interaction with individual molecules which accommodate to the local surface temperature transfers a net momentum to the particle (color online).}
\end{figure}

The basic idea of photophoresis in the free molecular flow regime is straightforward (Fig. \ref{fig:photophoresis}).  

An illuminated particle has a warm and a cold side. Individual molecules impact the surface. A fraction, $\alpha$, accommodates to the surface temperature and is  reejected at a velocity related to the surface temperature. This imparts a net momentum on the particle. Under typical conditions of protoplanetary disks the free molecular flow regime applies to dust size particles and the photophoretic force can be approximated by \citep{rohatschek1995}.

\begin{equation}
F_{Ph}=\frac{\alpha \pi a^3 I p}{6T} \cdot \frac{1}{k}
\label{photophorese}
\end{equation}

where $a$ is the particle radius, $I$ is the radiant flux density, $T$ is the gas temperature, $p$ is the gas pressure, $\alpha$ is the thermal accommodation coefficient (somewhat smaller than 1), and $k$ is the thermal conductivity of the particle.

In a minimum mass solar nebula the mean free path of the gas molecules in the midplane of the disk is 6 cm at 1 AU  \citep{HayashiEtal1985}. With the density increasing with decreasing 
radial distance as $R^{-11/4}$, the mean free path at 0.1 AU is still 60 $\mu m$. As we
consider photophoresis for micrometer grains they are well within the free molecular flow
regime and eq. \ref{photophorese} can be applied throughout the disk. 

As can be seen in eq. \ref{photophorese}, the thermal conductivity, $k$, is a major input parameter for photophoresis. Iron has a thermal conductivity larger than 50 W/(mK). Silicates have thermal conductivities on the order of 1 W/(mK). This is a difference of 
a factor 50. Therefore, metals are not influenced as strongly by photophoresis. Experiments in the drop tower Bremen by \citet{wurm2010}  showed this clearly as 1mm size silicate particles were accelerated upon illumination while 1mm steel particles did not show a measurable motion due to photophoresis. Aggregates of particles have lower thermal conductivities by 1 to 3 orders of magnitude \citep{presley1997, krause2011, vonBorstel2012}.
This is valid for aggregates consisting of silicates as well as aggregates consisting of metals.

\subsection{Metal and Silicate separation }

We assume that inner terrestrial planets form roughly at the location where they are currently found and grow by collecting mass in their feeding zone or gravitational reach. The question on compositional separation and selective growth therefore is if provided silicates are removed from a local feeding zone efficiently while metal grains stay to be accreted. The photophoretic force depends on the radiant flux density and is directed radially. If gravity is still dominating over photophoresis the primary effect of photophoresis would be to slow down the orbital velocity but particles would still be on a stable orbit. However, in a gaseous disk small sized dust particles couple to the gas motion on timescales much smaller than the orbital timescale. The consequence is that if the gas is (almost) on a Keplerian orbit and strongly photophoretically supported particles are slower, the particles speed up by gas friction and now move outward \citep{weidenschilling1977}. Photophoresis therefore can be treated as purely radial force for small particles moving particles outward \citep{wurm2006}. The drift velocity can be calculated as 

\begin{equation}
v=F/m\cdot \tau, 
\end{equation}

where $\tau$ is the gas grain coupling time. In the free molecular flow regime $\tau$ is given as \citep{blum1996} 

\begin{equation}
\tau = \gamma \frac{m}{\sigma} \frac{1}{\rho_g v_g}
\end{equation}

where $\gamma = 0.68$ is empirical, $m$ and $\sigma=\pi a^2$ are particle mass and cross section and $\rho_g$ and $v_g$ are
the gas density and average thermal velocity, respectively.
The coupling time depends inversely on density and therefore on pressure (ideal gas equation) 

\begin{equation}
\rho_g = \frac{p  \mu}{R_g T}
\end{equation}

where $R_g$ is the gas constant and $\mu = 2.34$ g/mol is the molar mass of the molecules.
As the photophoretic force depends linearly on pressure the drift velocity is independent of the gas pressure or density. In other words, if the gas density is lower, the photophoretic force is smaller but the particle has more time to speed up before the motion is damped. The terminal drift velocity is

\begin{equation}
v_{Ph} = \frac{\alpha \gamma R_g }{6 \mu} \frac{I}{v_g} \frac{a}{k}
\label{driftvelocity}
\end{equation}

It is therefore a convenient feature of photophoretic drift that it is not important what the detailed density profile of the disk is. 

At the position of Mercury (0.4 AU) $I = 8.5 \rm kW /m^2$. The temperature is about 600K which results in $v_g = 2300$ m/s. $\alpha$ is assumed to be 1. Therefore, for pure metal grains we get drift velocities of $v_{Ph} = 0.3 $ mm/s for a 
$10 \rm \mu m$ grain and for silicates we get $v_{Ph} = 15$ mm/s. 
Silicate particles set free in the inner system move out with a drift rate of $1 \rm AU$ in 300.000 years while metal grains require 50 times longer or 15 million years for 1 AU. This is also directly visible from eq. \ref{driftvelocity} which shows that the drift velocity is inversely depending on the thermal conductivity and the ratio of thermal conductivities between metal and silicate is 50. The absolute numbers show that the silicate grains on one side can keep up with an outward moving edge. On the other side metal particles are too slow to move significantly within the lifetime of a protoplanetary disk of a few million years. Therefore, if both species of dust particles are present, metal grains stay in the inner parts and silicate grains move outwards. Aggregates, even if they contain metals are pushed outward efficiently. Therefore, photophoresis provides a selection mechanism which is biased to pure bulk metal grains. This is the basic idea behind this paper. 

If individual particles are present at the optical thin but gaseous and dense inner edge of the disk then photophoresis selectively removes small particles that are not pure metal grains and the local assembly of large bodies is biased to metal rich material.

Two concerns have to be addressed further. (1) Can particles at the moving edge grow fast enough that planetesimals (and planets subsequently) form? (2) Are silicate and metal particles separated before or during this growth sufficiently that they are not incorporated into the same planetesimal?

\subsubsection{Formation of planetesimals}

It is sufficient to consider planetesimal formation as km-size precursors to planets here, not full planet formation as the small bodies will carry the sorting signature. Later stages of planet formation might then work via collisional growth and just assemble the sorted planetesimals \citep{wetherill1989, kokubo2010}. 

There are currently two more or less distinct ways to explain planetesimal formation. 
One way is to build planetesimals by collisions all the way from dust particles. Current numerical 
simulations by \citet{windmark2012} show that this is a viable way to produce larger bodies. 
It is important to note that fragmentation and therefore recycling of solids is an essential part of such models.
Even if growth of some particles proceeds the reservoir of small dust size particles is continuously replenished in collisions.
This has been shown in experiments by  \citet{wurm2005}, \citet{teiserwurm2009b} and \citet{kuepper2011}. 
Recent numerical simulations by \citet{garaud2012} show that two kind of dust populations coexist, small dust grains and
larger decimeter size bodies (depending on the radial distance).

The second way to form planetesimals is by instabilities in regions of high dust loading eventually leading to an enhancement of solids to a level that mutual gravitational attraction leads to the formation of planetesimals or even asteroids. This mechanism requires an essential part of the disk's solid mass to be in cm to dm size particles already. Assuming this is possible formation of a full size asteroid is found after only a few orbits \citep{johansen2011}. 

\citet{pinilla2012} explicitly call the edge of the disk a planetesimal factory. Eventually, photophoresis has to be plugged into such
codes to simulate the growth of sorted material.
However, these considerations alleviate the first concern mentioned above by demonstrating the capability of the edge to produce
planetesimals in time. 
To give some number here, let us assume -- as currently believed -- that due to inward drift and particle evolution there is a 
population of dm-size bodies and a population of dust grains. At a dust grain density of 32 $\rm m^{-3}$ as estimated above a larger
(decimeter) body of cross section $A$ moving through this cloud with velocity $v_d$ collects a volume of $A v_d$ per time or
$32 A v_d \rm m^{-3}$ particles per time. For $A = 0.01 \rm m^2$ and $v_d = 10 \rm m/s$ this is 3 $\rm s^{-1}$. As a ten micron particle has a volume
of about $10^{-15} m^3$ a decimeter body doubles in volume on a timescale of 10.000 years. This gives an estimate of the time scale
that small grains get incorporated into larger bodies which might then grow further or collapse gravitationally. This time scale is short compared
to the evolution of the disk edge. But as noted before,
there is a continuous recycling. Therefore, the larger (dm-size) bodies formed in later times will be more metal rich at the inward side of the edge  compared to the farther out optical thick end of the edge. On the timescale of 10.000 years
the silicate dust fraction already moved significantly by 10 \% of the edge thickness or 10.000 km. As the dust grains are collected by small bodies, separated particles will
not hit the same object but there is an average sorting during growth and fragmentation.

The assembly of still larger bodies from these decimeter objects is not a time problem as e.g. 
the gravitational instability model suggests that asteroid size
bodies can form within a few orbits from decimeter sizes. In total there is certainly enough mass as we do not consider a disk which is
depleted in gas and solids here. The disk is as dense as the initial disk at the edge only decreasing in density after the 
edge moved further out.  Depending on the mass fraction of small grains only the thickness of the edge varies. 
If a massive extrasolar planet or only a Mercury size planet can form depends on the detailed mass balance in the disk but
in principle there is no shortage of material. The sorting does not depend on the overall mass budget.

After separation the silicate particles have been pushed  to the outer edge of the edge of the protoplanetary disk by photophoresis where they blend into the bulk of unsorted particles. In total this will enrich the outer part in silicates.

\section{Conclusion}

The processes at an edge in a protoplanetary disk are complex. There are many aspects which influence the formation of larger bodies and planets eventually. Particle growth \citep{blumwurm2008}, radial drift \citep{weidenschilling1977}, and disk clearing 
\citep{alexander2007}, just to name a few. All these aspects are related. For some processes the timing is critical. The asteroids of the asteroid belt, e.g. are supposed to be formed up to 4 million years after the formation of the protoplanetary disk as chondrules supposed to be formed in the disk are dated to these different time scales \citep{Connelly2012}. Obviously, solid planetary bodies do not necessarily have to form right in the beginning, when the disk formed. Due to the photophoretic force particles at the inner edge of a disk are sorted with pure metal grains 
being enriched close to the inside of the edge as silicate particles move outward. This way high density planets like Mercury, Corot 7b, or Kepler 10b can be explained and the depletion of metals in asteroids is explained at the same time. However, as scattering processes in close encounters later on can also move objects from the metal rich region outward, metal rich enstatite parent bodies can be explained as well. Especially in view of recent observations of Mercury \citep{peplowski2011, nittler2011}, it might be useful to explore this formation or basic sorting mechanism in more detail as it is certainly a very favourable feature of photophoresis that it naturally connects a few different aspects of solar system and extrasolar planetary system composition.

\begin{acknowledgements}
Part of this work is based on DFG funded projects within the research group FOR 759 and the priority program SPP 1385.  
We also thank the referee for a thorough review of the manuscript.    
\end{acknowledgements}

\bibliographystyle{aa} 

\end{document}